\PassOptionsToPackage{unicode}{hyperref}
\PassOptionsToPackage{hyphens}{url}
\documentclass[
]{article}

\usepackage{xcolor}
\usepackage[margin=1in]{geometry}
\usepackage{amsmath,amssymb}
\usepackage{iftex}
\ifPDFTeX
  \usepackage[T1]{fontenc}
  \usepackage[utf8]{inputenc}
  \usepackage{textcomp} 
\else 
  \usepackage{unicode-math} 
  \defaultfontfeatures{Scale=MatchLowercase}
  \defaultfontfeatures[\rmfamily]{Ligatures=TeX,Scale=1}
\fi
\usepackage{lmodern}
\ifPDFTeX\else
\fi
\IfFileExists{upquote.sty}{\usepackage{upquote}}{}
\IfFileExists{microtype.sty}{
  \usepackage[]{microtype}
  \UseMicrotypeSet[protrusion]{basicmath} 
}{}
\usepackage{setspace}
\makeatletter
\@ifundefined{KOMAClassName}{
  \IfFileExists{parskip.sty}{%
    \usepackage{parskip}
  }{
    \setlength{\parindent}{0pt}
    \setlength{\parskip}{6pt plus 2pt minus 1pt}}
}{
  \KOMAoptions{parskip=half}}
\makeatother
\usepackage{graphicx}
\makeatletter
\newsavebox\pandoc@box
\newcommand*\pandocbounded[1]{
  \sbox\pandoc@box{#1}%
  \Gscale@div\@tempa{\textheight}{\dimexpr\ht\pandoc@box+\dp\pandoc@box\relax}%
  \Gscale@div\@tempb{\linewidth}{\wd\pandoc@box}%
  \ifdim\@tempb\p@<\@tempa\p@\let\@tempa\@tempb\fi
  \ifdim\@tempa\p@<\p@\scalebox{\@tempa}{\usebox\pandoc@box}%
  \else\usebox{\pandoc@box}%
  \fi%
}
\def\fps@figure{htbp}
\makeatother

\makeatletter
 \let\@cite@ofmt\@firstofone
 \def\@biblabel#1{}
 \def\@cite#1#2{{#1\if@tempswa , #2\fi}}
\makeatother
\newlength{\cslhangindent}
\newlength{\csllabelwidth}
 {\begin{list}{}{%
  \setlength{\itemindent}{0pt}
  \setlength{\leftmargin}{0pt}
  \setlength{\parsep}{0pt}
  \ifodd #1
   \setlength{\leftmargin}{\cslhangindent}
   \setlength{\itemindent}{-1\cslhangindent}
  \fi
  \setlength{\itemsep}{#2\baselineskip}}}
 {\end{list}}
\usepackage{calc}

\usepackage[hang,flushmargin]{footmisc}
\usepackage{booktabs}
\usepackage{subcaption}
\usepackage{bookmark}
\IfFileExists{xurl.sty}{\usepackage{xurl}}{} 
\hypersetup{
  hidelinks,
  pdfcreator={LaTeX via pandoc}}
\usepackage{ulem}

\usepackage{natbib}

\title{\textbf{Scattered spring: How climate change disrupts the synchrony of biological events }\\[1em]}

\author{
Jonathan Auerbach\footnotemark[1]\\
\textit{Department of Statistics} \\ 
\textit{George Mason University}\\
\textit{Fairfax, VA USA}\\
\and
Andrew Gelman\footnotemark[2]\\
\textit{Department of Statistics} \\ 
\textit{Department of Political Science}\\
\textit{Columbia University} \\
\textit{New York, NY USA}\\
\and
E. M. Wolkovich\footnotemark[3]\\
\textit{Forest and Conservation Sciences} \\
\textit{University of British Columbia}\\
\textit{Vancouver, BC Canada} \\[1em]
}

\date{(Dated: July 6, 2026)}

\begin{document}

\twocolumn[
\maketitle

\begin{quote}
Many biological processes, including plant leafout and flowering, occur once cumulative temperatures reach a threshold---a relationship known as the thermal-sum model. In this way, temperature is thought to coordinate the timing of biological events. An important implication is that higher temperatures cause thresholds to be reached sooner so that the timing of spring events, for example, should advance earlier in the calendar year as climates warm. But growing evidence has found that, as climates have warmed, the rate of advancement has slowed---a trend known as declining sensitivity---while the variance in the timing of spring events has increased in many cases---a trend we call declining synchrony. These trends raise questions about the resilience of temperature-based coordination to anthropogenic climate change. To answer these questions, researchers have modified the thermal-sum model by introducing additional factors and mechanisms, such as chilling and photoperiod. We show such complexity is not necessary to explain current trends of sensitivity and synchrony. Using experimental and real-world data, we find these trends are exactly as predicted by the thermal-sum model. In particular, the model predicts that as temperatures continue to increase and springtime events shift from the equinox toward the winter solstice, those events will become less synchronized and more variable, a phenomenon we refer to as a scattered spring. By 2100, our results predict much of the American South will experience a scattered spring under a high-warming scenario, with the average time between spring flowering events increasing by several weeks in many locations.
\end{quote}

\vspace{2em}
]
{\renewcommand{\thefootnote}{\fnsymbol{footnote}}
\footnotetext[1]{jauerba@gmu.edu}
\footnotetext[2]{ag389@columbia.edu}
\footnotetext[3]{wolkovic@mail.ubc.ca}

\section{Introduction}\label{introduction}

Many biological events are triggered not by temperature on a single day, but by temperature accumulated over many days \citep{larcher1980}. The long-standing and widespread `thermal-sum' model states that the bud of a flowering plant, for example, first blooms in the spring once cumulative temperatures exceed a plant-specific threshold \citep{quetelet1849letters, chuinearees}. This mechanism is thought to synchronize a wide variety of seasonal events within an ecosystem---for instance, ensuring that pollinators emerge when flowers are in bloom  \citep{schwartzbook}. It also supports an array of forecasting tools in which scientists track cumulative temperatures or `growing degree days' to predict harvest dates, manage pest emergence, and assess how changing temperature patterns reshape ecological communities \citep{schwartzbook}.

Shifts in the timing of biological events have been among the clearest fingerprints of anthropogenic climate change across both natural and managed systems \citep{ipcc2022}. But researchers have struggled to fully predict or explain these shifts in recent years. One puzzle is that while the timing of many biological events has moved earlier (i.e., advanced) as climates warm, the rate of advancement has slowed \citep{fu2015declining}. We refer to this pattern in which events appear less sensitive to additional warming as `declining sensitivity.'

A second puzzle is that the variance in the timing of biological events has in some cases declined and in others increased \citep{vitasse2018global,stemkovski2023disorder}. Higher variance has also manifested as previously unknown or rare events becoming more common \citep{chuine2025living}. We refer to this pattern in which events appear less coordinated as `declining synchrony.'

Declining sensitivity and synchrony are almost always interpreted as evidence that the long-standing and widespread thermal-sum model is wrong or incomplete. Researchers have suggested myriad complexities or alternatives \citep{fu2015declining,wolkovich2021simple,gao2024thermal}. Yet no work has examined whether these puzzles are explained by the thermal-sum model, or how sensitivity and synchrony will continue to evolve under the thermal-sum model as climates continue to warm.

Here we show that the thermal-sum model is an example of a stopped random walk, and we build a simple and transparent framework for the analysis of biological event times using statistical tools developed for stopped random walks. Applying the framework to observational and experimental data, we find that the thermal-sum model explains both declining sensitivity and synchrony. Thus modifications and alternative approaches often suggested  \citep{fu2015declining,vitasse2018global} may be unnecessary and yield inaccurate predictions. In contrast to predictions from current models of biological events, our framework implies that as springtime events advance with anthropogenic warming, they enter a new temperature regime in which the typical time between events in the same ecosystem can increase from a few days to a month or more, a phenomenon we refer to as a `scattered spring.'

\section{Results \& Discussion}\label{resultsdiscuss}

The thermal-sum model states that a biological event occurs when cumulative temperature reaches a threshold. This model is an example of a stopped random walk where the cumulative temperature is the position of the walk, and the biological event occurs when the walk arrives at the threshold. 

Consider the timing of a specific biological event, such as the bloom date of a bud. Let $X_i$ denote the effective temperature experienced by that bud on day $i$. We model $X_i$ under two temperature regimes (Figure \ref{fig:model}).

\begin{figure*}[t]
\centering
\pandocbounded{\includegraphics[keepaspectratio]{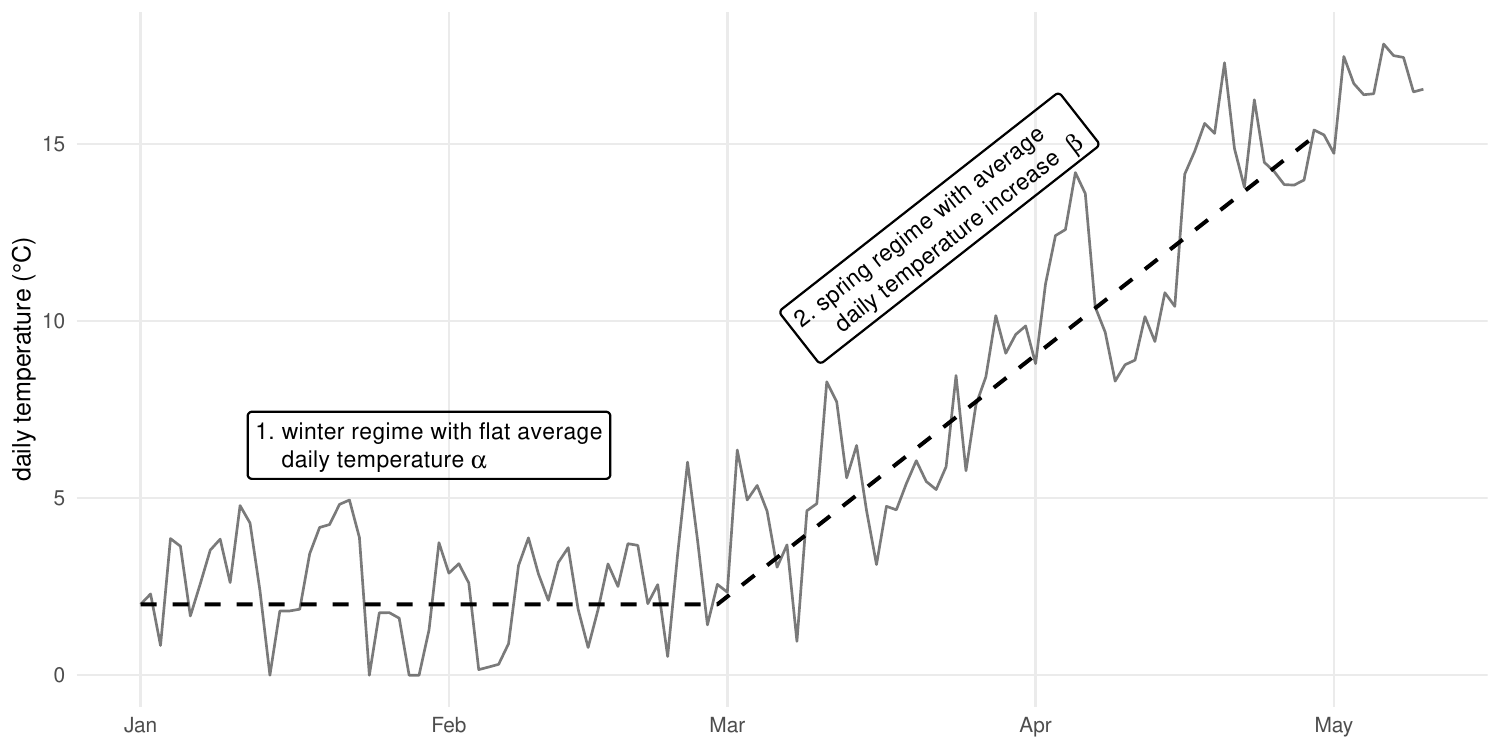}} 
\caption{We model the daily temperature (solid line) as trend (dashed line) plus noise. The timing of springtime events in temperate climates depends on whether temperature is accumulated during one of two regimes, 1. the winter regime (relatively flat average daily temperature near the winter solstice, which we refer to as the `winter temperature' or $\alpha$), or 2. the spring regime (relatively linear average daily temperatures near the spring equinox, the slope of which we refer to as `spring warming' or $\beta$). We assume the effective temperature experienced by the bud is positive since biological limits prevent plants from accumulating very low forcing temperatures, taken here to be $0^{\circ}$C \citep{larcher1980} although this is easily adjusted.}
\label{fig:model}
\end{figure*}

The first regime is the `winter regime' in which the average daily temperature is approximately constant at some level $\alpha>0$. This regime is often observed in the winter months following the solstice, and we refer to $\alpha$ as the `winter temperature.'  The second regime is the `spring regime' in which the average daily temperature rises at a constant rate $\beta>0$. This regime is often observed in the spring months closer to the equinox, and we refer to $\beta$ as `spring warming.' 

In either regime, the effective temperature is modeled as the average temperature plus noise $\epsilon_i$. That is, $X_i = \alpha + \beta i + \epsilon_i,$ where $\beta=0$ in the winter regime and $\beta>0$ in the spring regime. We assume the $\epsilon_i$ are independent with mean $0$ and variance $\sigma^2$.

Noise in this formulation can represent the influence of various biological factors, depending on the scale of analysis and on our evolving understanding of what triggers springtime events. At the bud level, it represents idiosyncratic variation specific to that bud, which could arise from microclimate differences \citep{schwartz2014separating}, bud color variation \citep{peaucelle2022accurate}, or internal biological differences that affect when the bud begins accumulating temperatures \citep{vanderschoot2014,pan2023epigenetic}.

This two regime model in which average temperatures switch from constant to increasing well-describes both historic springtime temperature data (Figure \ref{fig:temp_plot-1} in SI.3) and projected springtime temperature data (Figure \ref{fig:temp_plot-2} in SI.3). Here we assume that warming occurs entirely in one of the two regimes, although similar results hold when warming occurs under both regimes (since in such cases the spring regime dominates, see SI.1 for details.)

The bloom date $\nu(\tau)$ is the day $n$ on which the cumulative temperature $\sum_{i=1}^n X_i$ first reaches a plant-specific threshold $\tau>0$. This formulation is standard in growing degree day models and, more generally, the `forcing' component of many process-based models of plant phenology \citep{schwartzbook}. This formulation also allows us to use standard results for stopped random walks, which yield asymptotically normal approximations for the bloom date when the threshold is sufficiently large that accumulation occurs over many days.

In the winter regime, average temperatures are constant during accumulation ($\beta = 0$), and cumulative average temperatures grow at an approximately linear rate. The average bloom date advances with climate change in step with the winter temperature ($\mathbb{E}[\nu(\tau)] \approx \tau / \alpha$) as is often described in the literature \citep{vitasse2022great}. In the spring regime, average temperatures increase during accumulation ($\beta > 0$), and cumulative average temperatures grow quadratically. In this regime, the average bloom date squared advances in step with spring warming ($\mathbb{E}[\nu(\tau)] \approx \sqrt{ 2 \tau / \beta}$). 

Within either regime, the relationship between the average bloom date and temperature is non-linear as observed in recent papers \citep{fu2015declining,wolkovich2021simple}. Indeed, both regimes predict declining sensitivity in that the average bloom date advances at a diminishing rate when warming increases as measured by the winter temperature ($\alpha$) or spring warming ($\beta$).

Importantly, the effects of warming within the two regimes make very different predictions regarding the synchrony of events---that is, the variance in the timing of events within the same year and location due to idiosyncratic differences described above such as microclimate. Both predict decreased variance with warming ($\mathrm{Var}(\nu(\tau)) \approx \sigma^2 \, \tau / \alpha^3$ in the winter regime and $\mathrm{Var}(\nu(\tau)) \approx \sigma^2 / \sqrt{2 \beta^{3} \tau}$ in the spring regime). But the overall size of the variance depends on whether the threshold $(\tau)$ is in the numerator or the denominator.

When the threshold is in the denominator as in the spring regime, the variance is small, reflecting the fact that as temperatures accumulate, noise is quickly overwhelmed by the signal (quadratic increase in cumulative average temperatures). This is how temperatures near the equinox synchronize plants (and buds within plants) with the same threshold---by making idiosyncratic differences such as microclimate negligible. In contrast, when the threshold is in the numerator as in the winter regime, the variance is large, reflecting the fact that noise remains relatively influential (compared to the linear increase in cumulative average temperatures) and idiosyncratic differences translate into larger differences in threshold-crossing times. 

In this way, our results reveal the key mechanism that determines whether synchrony declines: the temperature regime under which accumulation occurs. When average temperatures increase day over day, as they often do near the equinox, springtime events tend to be tightly synchronized. But when warming shifts those events closer to the solstice, where average temperatures are comparatively constant, event times become more variable. The shift in regimes is observable as a `scattered spring' in which, as springtime events advance, they become less synchronized. 

Our results show that the two most important aspects of temperature are the levels of winter temperature $(\alpha)$ and spring warming $(\beta)$. This simplification yields a simple and transparent framework for translating spatially varying climate change into spatially varying predictions of both the advancement of spring and its variability. 

To demonstrate the framework, we analyze bloom dates from the lilac--honeysuckle phenology network, compiled by \cite{rosemartin2015lilac} from sites across the continental United States between 1956--2014. We limit our analysis to the purple common lilac (\textit{Syringa vulgaris}) and the phenophase full bloom, and we update the dataset using USA National Phenology Network records from the same sites and species until 2025 \citep{switzer_etall:2025}. For each record, we estimate the winter temperature $(\alpha)$ and spring warming $(\beta)$. We then estimate the conditional mean and standard deviation functions of the bloom date using a generalized additive model \citep{hastie1986generalized, wood2017generalized}. 

We find that increasing either the winter temperature, spring warming, or both has shifted lilac flowering times exactly as predicted by the thermal-sum model (Figures \ref{fig:lilac_te}). Warming in either regime (increasing $\alpha$ or $\beta$) leads to earlier mean bloom dates, which advance at a decreasing rate (declining sensitivity). Warming in the winter regime only (increasing $\alpha$ and holding $\beta$ fixed) returns greater variance (higher standard deviation, declining synchrony), while warming in the spring regime only (increasing $\beta$ while holding $\alpha$ fixed) returns lower variance (lower standard deviation, increasing synchrony). This is the `scattered spring' predicted by the model: in the first case, plants are more likely to accumulate their thermal sum in the winter regime, which results in the larger variance.

\begin{figure*}
    \centering

    \begin{subfigure}[b]{0.9\textwidth}
        \centering
        \includegraphics[width=1\linewidth]{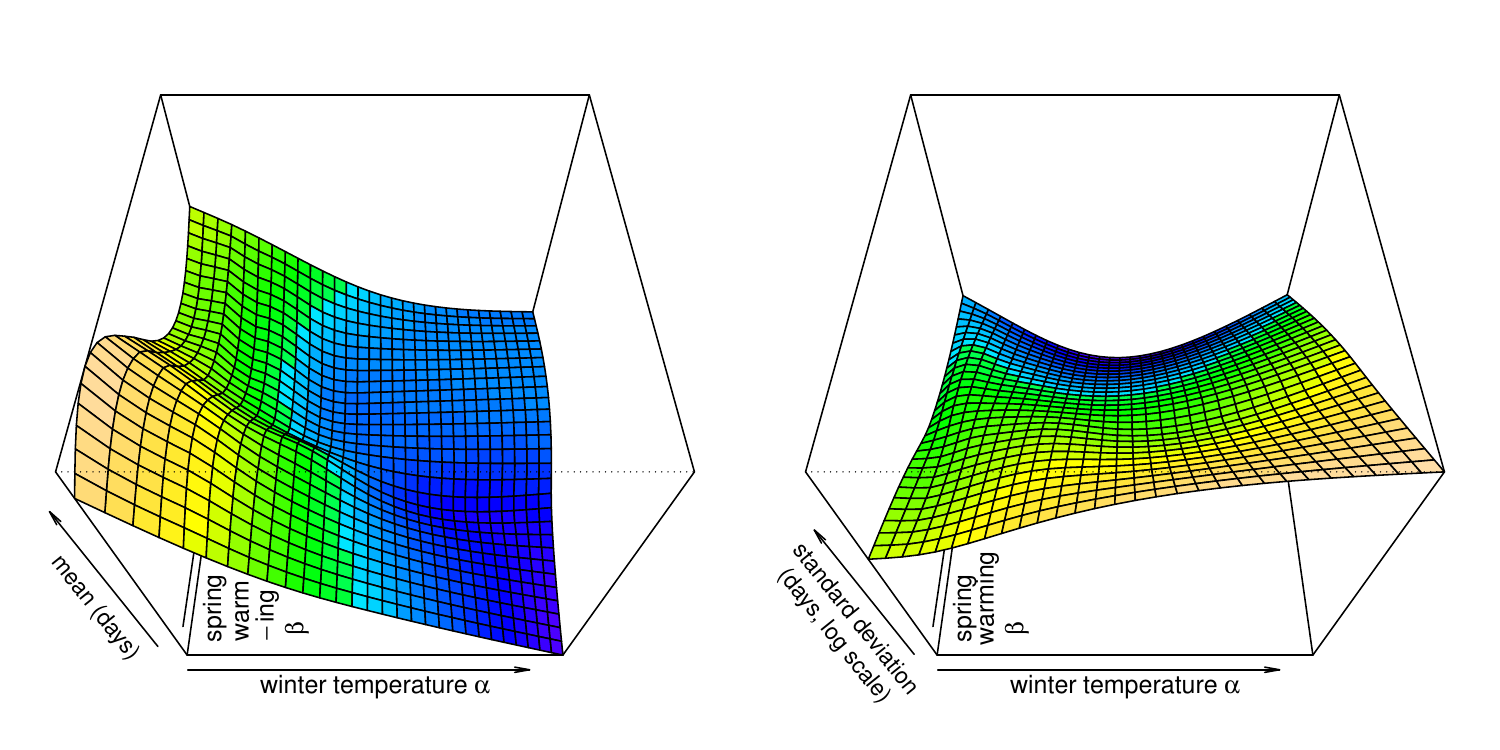}
        \caption{Bivariate plots showing mean (left) and log standard deviation (right) of lilac bloom dates as a function of winter temperature (\(\alpha\)) and spring warming (\(\beta \)).}
    \end{subfigure}

    \vspace{0.5em}

    \begin{subfigure}[b]{0.9\textwidth}
        \centering
                \includegraphics[width=1\textwidth, trim=0 0 .85in 0, clip]{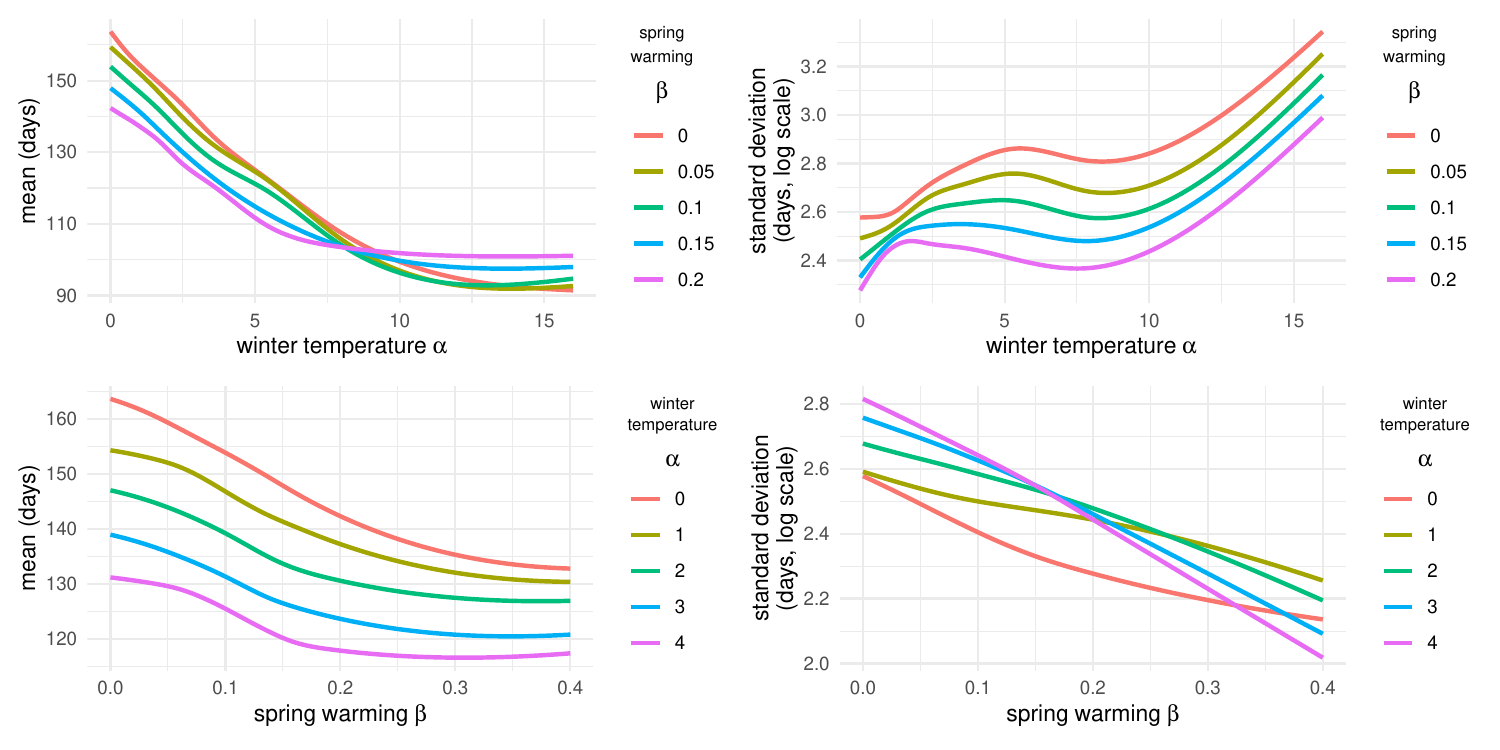}
        \caption{Cross-section plots showing mean (left) and log standard deviation (right) of lilac bloom dates as a function of winter temperature (\(\alpha\)) holding spring warming (\(\beta \)) constant (top) and as a function of spring warming (\(\beta \)) holding winter temperature (\(\alpha\)) constant (bottom).}
    \end{subfigure}

     \caption{Mean and log standard deviation of lilac bloom dates (n = 10,613 from Rosemartin et al. 2015, updated to 2025 using USA National Phenology Network records) by winter temperature $(\alpha)$ and spring warming $(\beta)$ as estimated by a generalized additive model. The estimated functions are shown using bivariate plots (a, top) and cross-section plots (b, bottom). The bivariate plots show the mean decreases as $\alpha$ and $\beta$ increase (left), while the standard deviation can increase or decrease depending on the relative values of $\alpha$ and $\beta$ (right). The cross-section plots show the mean decreases both as \(\alpha\) increases holding \(\beta\) constant (top left) and as \(\beta\) increases holding \(\alpha\) constant (bottom left), while the standard deviation increase when \(\alpha\) increases holding \(\beta\) constant (top right) and decreases when \(\beta\) increases holding \(\alpha\) constant (bottom right). Temperature is measured in $^\circ$C.}
    \label{fig:lilac_te}
\end{figure*}
 
In this way, the thermal-sum model explains a wide variety of trends in the mean and variance of biological event times, challenging a growing body of work claiming such trends either invalidate the model or justify additional complexity \citep[e.g.,][]{fu2015declining,kovaleskipreprint}. For example, researchers have suggested that climate change may influence photoperiod or chilling (cool temperatures required to break endodormancy), which in turn change the threshold of the thermal sum. 

We show that these additional factors are unnecessary by applying our framework to rare experimental data in which photoperiod and chilling are held fixed but forcing temperatures vary. We use results from \cite{Charrier:2011aa}, who tested the effect of different constant temperatures on walnut (\textit{Juglans regia} and \textit{Juglans regia x nigra} hybrids) tree budburst after the plants had experienced sufficient chilling to break endodormancy (385 buds were observed on stems sampled from 15 trees, chilled at \(4^\circ\)C), and we fit the winter regime approximation to the thermal-sum model (since temperatures are held constant, this experiment replicates this regime) using weighted least squares. The data exhibit the exact nonlinear relationship in the mean and variance as predicted by the thermal-sum model (Figure \ref{fig:experiment}), establishing the sufficiency of the model even under an extreme temperature range and controlled chilling and photoperiod conditions. 

\begin{figure*}[t]
\includegraphics[width=1\linewidth]{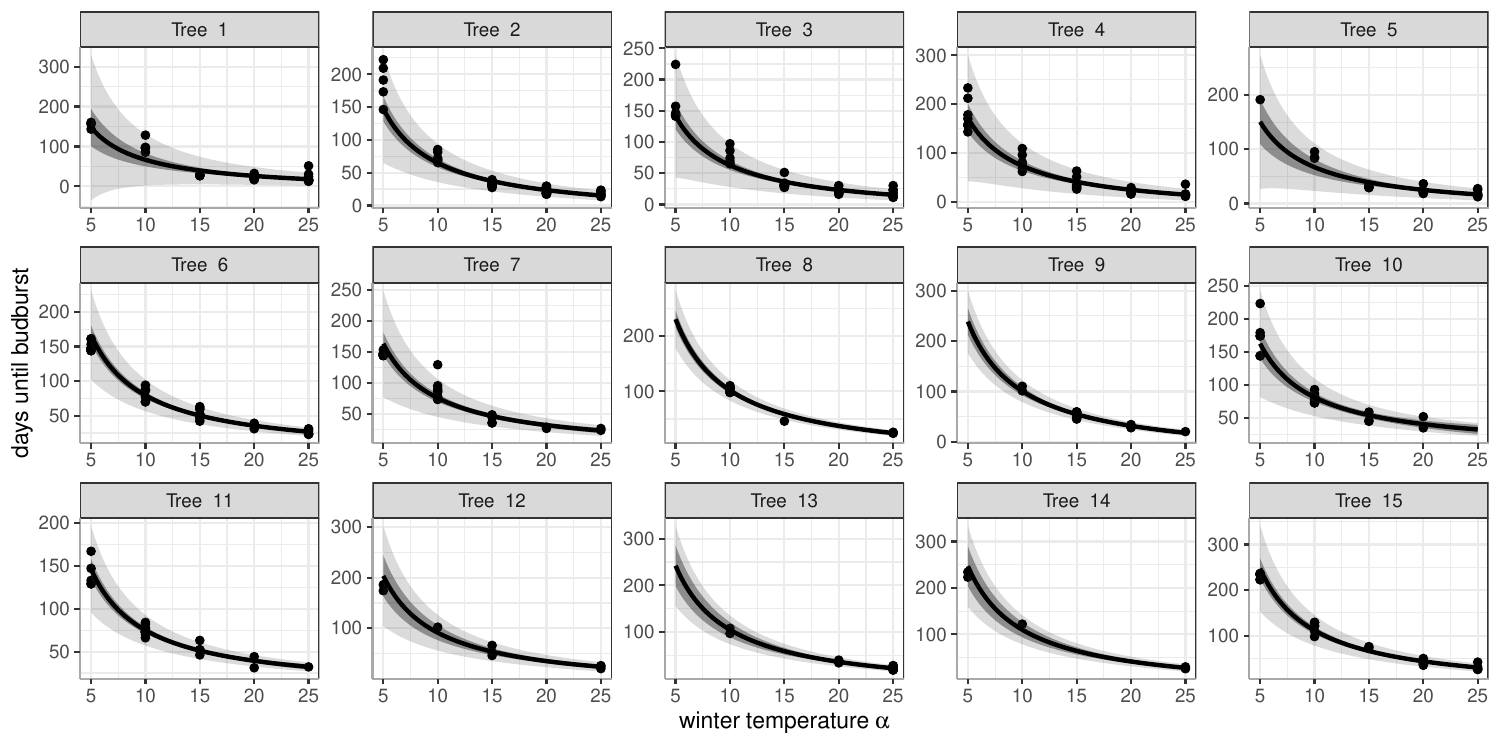} \caption{Mean time until budburst as a function of winter temperature ($\alpha$) in a controlled experiment conducted by Charrier et al. (2011) in which 385 buds were observed on stems sampled from 15 walnut trees (\textit{Juglans regia} and \textit{Juglans regia x nigra} hybrids), chilled at \(4^\circ\)C, and then warmed in different temperatures environments (x-axis) with the same photoperiod conditions across temperatures until budburst. For each tree (panel), we show the observed dates (points) and the fitted winter regime approximation of the thermal-sum model (the solid line denotes the mean, the inner grey region denotes the 95\% confidence interval of approximately two standard errors, and the outer grey region denotes the 95\% prediction interval of approximately two standard deviations). The figure highlights that the mean and variance are nonlinear functions of temperature even when biological factors such as chilling and photoperiod are held fixed. Both decrease with increasing temperatures as predicted by the thermal-sum model in the winter regime.} 
\label{fig:experiment}
\end{figure*}

Our results show that the long-standing and widely-used thermal-sum model \citep{quetelet1849letters, chuinearees} explains observed shifts in the timing of biological events including declining sensitivity and synchrony. This finding has two important consequences. First, it challenges an extensive body of work that has attempted to revise the thermal-sum model by adding additional factors or mechanisms \citep{fu2015declining}. These factors can influence the timing of biological events, but our work suggests they do not themselves explain current trends in sensitivity and synchrony. Further, additional factors can produce inaccurate forecasts when modeled incorrectly.

To understand the role played by any additional factors, the basic relationship between event time and temperature must be modeled correctly. For example, the literature often assumes a linear relationship between temperature and event time with constant variance, which we have shown is inconsistent with the thermal-sum model. Our framework provides a simple and transparent way to encode the thermal-sum model, on top of which complexity may be added and better studied.

The second consequence is that the thermal-sum model predicts a `scattered spring' or general desynchronization of springtime events as their timing moves closer to the winter solstice. We find that the synchrony of lilac bloom dates, a common indicator of spring, will decline in much of the Southern United States by the year 2100 under a high-warming scenario (SSP5-8.5): For example, were lilac planted across the South in 2050, the average time between blooms would be less than two weeks in 76\% of locations. By 2100, however, the average time will be less than two weeks in only 18\% of locations (Figure \ref{fig:forecast_south}, note that the synchrony of species with lower threshold $(\tau)$ may decline sooner). In many locations, the average time will increase by several weeks.

Absent some compensating mechanism, the long-standing and widespread thermal-sum model predicts that the timing of biological events across ecosystems---from crops to forests and from plants to insects---will in many cases become more variable, with potential cascading consequences for industry and natural ecosystems.

\begin{figure*}
    \centering

    \begin{subfigure}[b]{0.9\textwidth}
        \centering
	   \includegraphics[width=1\linewidth]{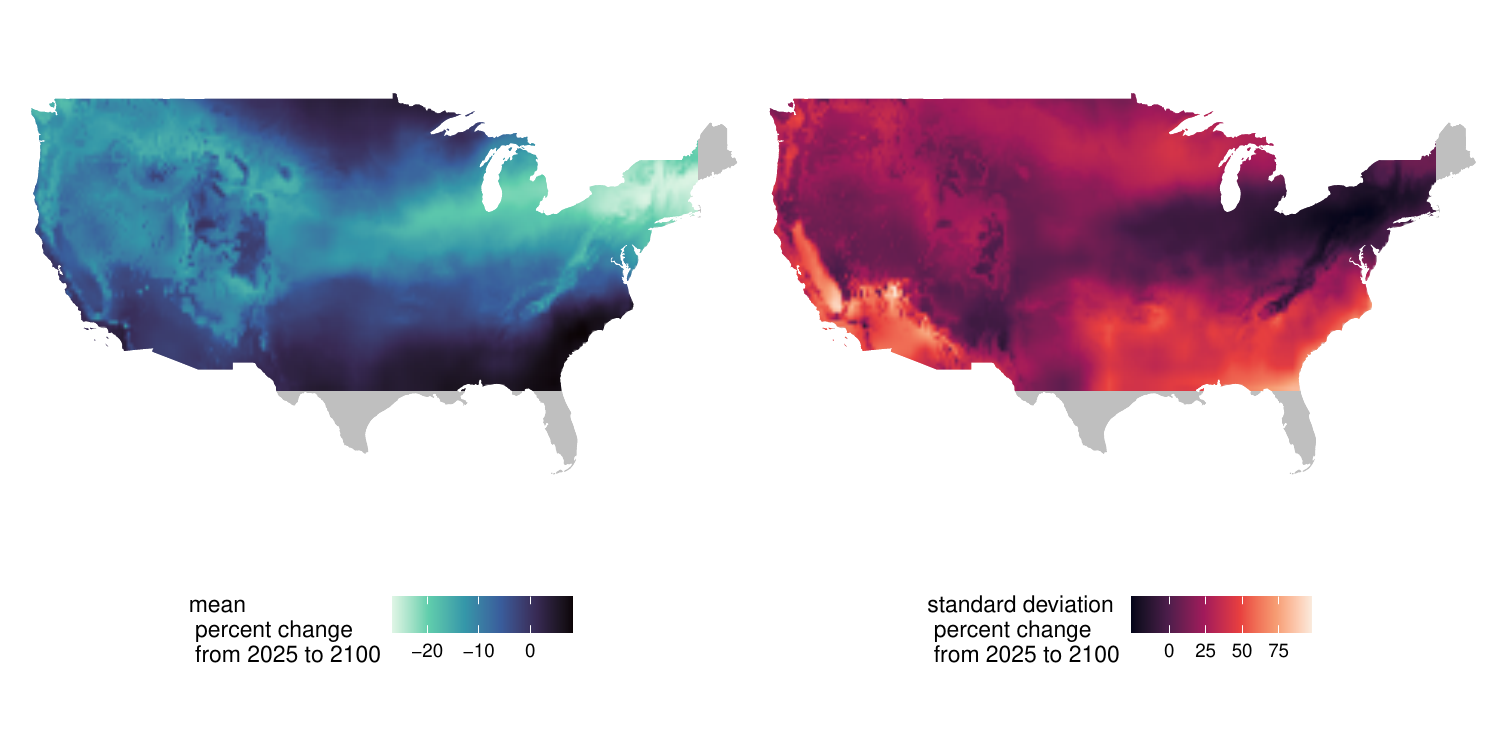}
    \end{subfigure}

     \caption{Percent change in the mean (left) and standard deviation (right) of lilac bloom dates between the years 2025 and 2100 across the continental United States under temperatures projected in the NEX-GDDP-CMIP6  high-warming scenario SSP5-8.5 \citep{thrasher2022nasa}. Lighter areas indicate larger changes. The mean bloom date advances moderately, with the average bloom date occurring 20\% earlier in parts of the Northeast. In contrast, the standard deviation increases dramatically in some areas, with the average time between events increasing by 100\% in parts of the South, a phenomenon we refer to as a scattered spring.}
    \label{fig:forecast_south}
\end{figure*}

\subsection{Methods}\label{methods}

Our observational data analysis considers all sites in the historical lilac--honeysuckle phenology network \citep{rosemartin2015lilac} that recorded the phenophase full bloom and were within 10 miles of a Global Historical Climatology Network daily (GHCNd)  site with sufficiently complete temperature records. We limit our analysis to common lilac (\textit{Syringa vulgaris}), including all observations made from 1955 to 2025 as recorded by the USA National Phenology Network. The data were retrieved using the \texttt{R} package \texttt{rnpn}, resulting in \(10,613\) observations. 

For each observation, we estimate the winter temperature \((\alpha)\) using the average daily temperature between January and February, and we estimate spring warming \((\beta)\) using the slope of a linear regression model fit by regressing daily temperature on day index (i.e., $i=1$ for January 1, $i=2$ for January 2, etc.) over March and April (Figure \ref{fig:temp_plot-1}, \ref{fig:temp_alpha}, and \ref{fig:temp_beta} in SI.3). Temperatures below \(0\) are set to \(0\), a common base temperature in growing degree day models. 

We then fit a two-parameter Gaussian location--scale model using the \texttt{gaulss} function from the \texttt{R} package \texttt{mgcv} in which we define 
\begin{align*}
 \mu(\alpha,\beta)&=\mathbb{E}[\nu(\tau) \mid \alpha,\beta] \\[1em]
 \sigma(\alpha,\beta)&=\mathrm{SD}(\nu(\tau) \mid \alpha,\beta).
 \end{align*} 
 Both \(\mu\) and \(\sigma\) are modeled with tensor-product splines with basis dimension k = 10 (Figure \ref{fig:lilac_te}). 
 
Finally, we calculate winter temperature $(\tilde \alpha)$ and spring warming $(\tilde \beta)$ for the years 2025, 2050, 2075, and 2100 using the NEX-GDDP-CMIP6 temperature projections for two scenarios: SSP2-4.5 and SSP5-8.5 \citep{thrasher2022nasa} (Figure \ref{fig:temp_plot-2} in SI.3). We also calculate the projected mean bloom date $ \mu(\tilde \alpha,\tilde \beta)$ and standard deviation $\sigma(\tilde \alpha, \tilde \beta)$ (Figure \ref{fig:forecast_south} and Figures \ref{fig:forecast_mean} and \ref{fig:forecast_var} in SI.3).

\section{Acknowledgments}

We thank G. Charrier for sharing data and V. Van der Meersch for helpful comments and suggestions.

\bibliographystyle{apalike}
\bibliography{auerbach_gelman_wolkovich}

\onecolumn

\section{S. Supporting Information}\label{s.-appendix}

\setcounter{figure}{0}
\renewcommand{\thefigure}{S\arabic{figure}}

\setcounter{table}{0}
\renewcommand{\thetable}{S\arabic{table}}

\subsection{SI.1 Sketch of Theoretical
Results}\label{s.1-sketch-of-theoretical-results}

We follow the notation in \cite{gut2009stopped}. Let
\(X_i=\alpha+\beta i + \epsilon_i\) denote the temperature on day \(i\),
where \(\alpha>0\), \(\beta\ge 0\), and the \( \{\epsilon_i\} \) are 
independent and identically distributed with  
\(\mathbb{E}[\epsilon_i]=0\) and 
\(\mathrm{Var}(\epsilon_i)=\sigma^2\). Let \(\gamma=\frac{\alpha}{\beta}+\frac{1}{2}\), and
define the deterministic
cumulative sum \[
\xi_n=\sum_{i=1}^n \mathbb{E}[X_i]
=
\alpha n+\frac{\beta}{2}n(n+1)
=
\frac{\beta}{2}n^2+\beta\gamma n\]
when \(\beta>0\) and \(\xi_n=\alpha n\) when \(\beta=0\). 

Denote the sum of the
stochastic component \(S_n=\sum_{i=1}^n \epsilon_i\). The object of
interest is the first-passage time of \( Z_n=S_n+\xi_n\), \[
\nu(\tau)=\min\{n\ge 1:\ Z_n> \tau\}.
\] 

In the language of \cite{lai1977nonlinear}, \(\{Z_n\}\) is a
perturbed random walk. That is, the random walk \(\{S_n\}\) is perturbed by the
deterministic term \(\{\xi_n\}\). This setup yields two
asymptotic regimes, corresponding to whether \(\beta\approx 0\) as in the winter regime or 
\(\beta>0\) as in the spring regime.

When \(\beta=0\), we apply the usual asymptotic argument for
the hitting time of a random walk with constant positive drift as \(\tau \to \infty\), \[
\nu(\tau)\ \dot{\sim}\ \mathrm{Normal}\!\left(\frac{\tau}{\alpha},\ \frac{\sigma^2\,\tau}{\alpha^{3}}\right),
\] see \cite{gut2009stopped}.

When \(\beta>0\), the deterministic component
\(\xi_n=\frac{\beta}{2}n^2+\beta\gamma n\) grows quadratically so that
the threshold \(\tau\) is reached near the deterministic crossing time
\(m(\tau)\) satisfying \(\xi_{m(\tau)}\approx \tau\). The solution is \[
m(\tau)=\frac{-\beta\gamma+\sqrt{\beta^2\gamma^2+2\beta \tau}}{\beta}
=
\sqrt{\frac{2\tau}{\beta}}-\gamma+o(1).
\] By the functional central limit theorem, the deviation of \(\nu(\tau)\)
about \(m(\tau)\) is approximately normal. Linearization of
\(Z_{\nu(\tau)}\approx \tau\) around \(m(\tau)\) yields \[
\nu(\tau)\ \dot{\sim}\ \mathrm{Normal}\!\left(m(\tau),\ \frac{\sigma^2\,m(\tau)}{\bigl(\alpha+\beta\,m(\tau)\bigr)^2}\right)
\;\approx\;
\mathrm{Normal}\!\left(\sqrt{\frac{2\tau}{\beta}}-\gamma,\ \frac{\sigma^2}{\beta^{3/2}\sqrt{2\tau}}\right),
\] where we have used the fact that
\(\alpha+\beta m(\tau) \approx \sqrt{2\beta \tau}\) when
\(\tau\) is large. 

We make two observations about this asymptotic approximation. First, we have written the cumulative sum as beginning on a known day \(i = 1\). 
However, the approximation does not actually depend on the day the accumulation begins. This is because when \(\tau\) is sufficiently large, the first finite number of terms of the cumulative sum are negligible by comparison. For example, for any start day \(a > 1\), incorrectly starting at day \(i = 1\) yields error \[\sum_{i=1}^{\nu(\tau)}X_i-\sum_{i=a}^{\nu(\tau)}X_i = \sum_{i=1}^{a-1}X_i = O_p(1).\]

Second, we have described accumulation as occurring entirely within one of two regimes. In practice, accumulation may begin near the end of the winter regime (\(\beta = 0 \)) 
and continue into the spring regime (\(\beta > 0\)). In such cases, the event time will behave as occurring in the spring regime when \(\tau\) is large as long as a nontrivial 
amount of warming happens under  the spring regime (i.e., the percent of time in the spring regime does not go to 0 as \(\tau\) goes to infinity). This is because the 
quadratic accumulation in the spring regime dominates the contribution from the winter regime.  Moreover, the linearization argument used to compute the 
variance shows that the asymptotic variance depends only on the accumulation that occurs near the event time. That is, the variance only depends on 
the regime active when the event occurs.

\subsection{SI.2 Additional Tables}\label{s.2-additional-tables}

In addition to the model-based approach in the main text, we also examine the data using a binning approach.  
Table \ref{tab:walnut_forcing} shows data from a controlled experiment on walnut trees
(\textit{Juglans regia} and \textit{Juglans regia x nigra} hybrids) conducted by \cite{Charrier:2011aa}. 
See also \cite{guillaume2018assessing}. The data reflect 385 buds from 15 trees comprising 6 genotypes at 2
locations. In November of the study year, stems were sampled from each
tree and cut into 7 centimeter segments containing a single bud. All
segments were chilled at \(4^\circ\)C and then transferred to constant
`forcing' environments with the air temperature set at one of \(5,10,15,20,\)
or \(25^\circ\)C, denoted \(\alpha\). For each temperature, the outcome is the response
time, the time (in days) until budburst determined from the date at
which the buds reached stage 10 of the BBCH scale.

\begin{table*}[t]
\centering
\caption{Summary data from the Walnut forcing experiment (Charrier et al. 2011).}
\label{tab:walnut_forcing}
\begin{tabular}{rrrr}
\toprule
$\alpha$ & $n$ & mean & sd \\
\midrule
 5  & 69 & 172 & 34.6  \\
10  & 86 &  88 & 17.1  \\
15  & 76 &  42 & 13.2  \\
20  & 77 &  26 &  8.4  \\
25  & 77 &  21 &  7.4  \\
\bottomrule
\end{tabular}

\vspace{2em}

\centering
\caption{Mean bloom date (day-of-year) of lilac (Rosemartin et al. 2015, updated to 2025 using USA NPN records) by winter temperatures ($\alpha$, rows) and spring warming ($\beta$, columns).}
\label{tab:mean_bloom_bins}
\begin{tabular}{lrrrr}
\toprule
$\alpha$ $\backslash$ $\beta$
& $[-0.28,\,0.07]$ & $(0.07,\,0.11]$ & $(0.11,\,0.15]$ & $(0.15,\,0.68]$ \\
\midrule
$(0,\,0.7]$     & 157.70 & 152.07 & 147.60 & 142.20 \\
$(0.7,\,1.9]$   & 151.12 & 146.88 & 141.64 & 136.15 \\
$(1.9,\,4.7]$     & 136.30 & 131.92 & 127.58 & 124.94 \\
$(4.7,\,16.4]$    & 109.56 & 109.77 & 107.36 & 105.71 \\
\bottomrule
\end{tabular}

\vspace{2em}

\centering
\caption{Standard deviation of lilac bloom date (days, Rosemartin et al. 2015, updated to 2025 using USA NPN records) by winter temperatures ($\alpha$, rows) and spring warming ($\beta$, columns).}
\label{tab:sd_bloom_bins}
\begin{tabular}{lrrrr}
\toprule
$\alpha$ $\backslash$ $\beta$
& $[-0.28,\,0.07]$ & $(0.07,\,0.11]$ & $(0.11,\,0.15]$ & $(0.15,\,0.68]$ \\
\midrule
$(0,\,0.7]$     & 12.80 & 11.94 & 10.90 &  10.95 \\
$(0.7,\,1.9]$   & 12.76 & 11.91 & 12.67 & 12.99 \\
$(1.9,\,4.7]$     & 16.68 & 14.97 & 14.45 & 12.55 \\
$(4.7,\,16.4]$    & 19.39 & 17.06 & 15.31 & 12.35 \\
\bottomrule
\end{tabular}
\end{table*}

This experiment isolates the winter regime because the forcing temperature is held
constant at level \(\alpha\) over time. To show consistency with this regime, Table 
\ref{tab:walnut_forcing} reports the mean and standard deviation
of the response times across the 30 stems at each forcing temperature.
Higher forcing temperatures advance budburst at a decreasing rate, 
and the standard deviation of budburst timing is 
smaller at warmer temperatures. Thus, in an experimental setting, the 
thermal-sum model captures the relationship between
constant-temperature forcing and the level and variability of
budburst times. This pattern cannot be explained by factors such as chilling or photoperiod, which are held constant.

Tables \ref{tab:mean_bloom_bins}-\ref{tab:sd_bloom_bins} show data from
the historical lilac--honeysuckle phenology network described in the main text. 
We bin sites and years within a grid of winter temperature (\(\alpha\), rows) and spring warming (\(\beta\), columns) 
and calculate the mean and standard deviation of the observed bloom date, as the number of days since January 1. 
The results are consistent with Figure \ref{fig:lilac_te} in the main text.
The mean and standard deviation are nonlinear functions of $\alpha$ and $\beta$. In particular, the standard deviation generally increases when
$\alpha$ increases (holding $\beta$ fixed) and decreases when $\beta$ decreases (holding $\alpha$ fixed). 

\subsection{SI.3 Additional Figures}\label{s.3-additional-figures}

The winter temperature (\(\alpha\)) and spring warming (\(\beta\)) are shown for nine sites in Figures \ref{fig:temp_plot-1}-\ref{fig:temp_beta}, chosen from the historical lilac--honeysuckle phenology network. Figure \ref{fig:temp_plot-1} shows nine randomly selected site-years and has the same interpretation as Figure 1. Figures \ref{fig:temp_alpha} and \ref{fig:temp_beta} show \(\alpha\) and \(\beta\) for the nine sites with the longest running records. Figure \ref{fig:temp_plot-2} shows \(\alpha\) and \(\beta\) for two random sites in the years 2050, 2075, and 2100 under temperature projections from NEX-GDDP-CMIP6 scenarios of moderate warming (SSP2-4.5) and high warming (SSP5-8.5) \citep{thrasher2022nasa}. Figures \ref{fig:forecast_mean} and \ref{fig:forecast_var} show the projected mean and standard deviation of lilac bloom dates using the \(\alpha\) and \(\beta\) calculated from the NEX-GDDP-CMIP6 data and the generalized additive model described in the main text. 

\vfill

\begin{figure*}
\centering
\pandocbounded{\includegraphics[keepaspectratio]{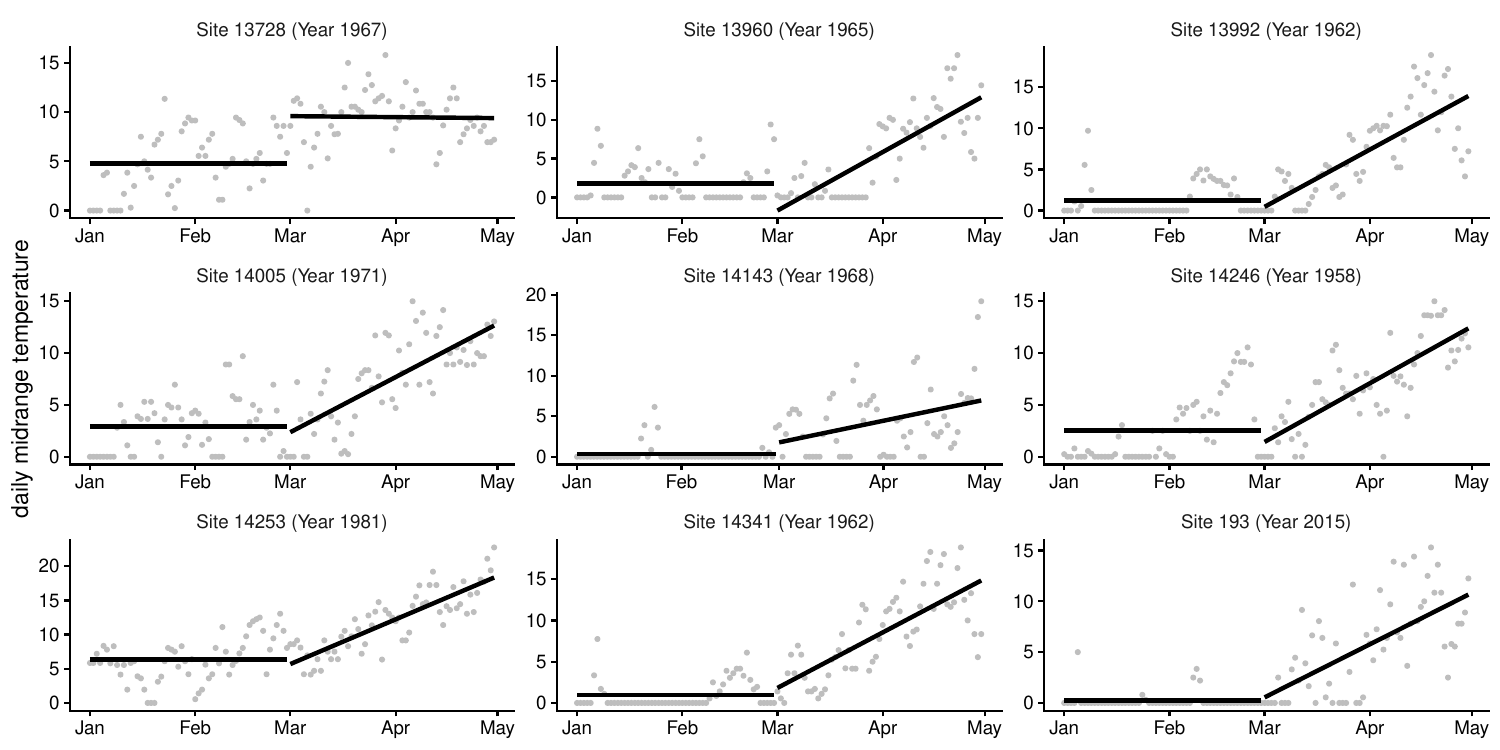}}
\caption{Springtime daily midrange temperatures (points) with \(\alpha\)
(mean from Jan to Feb) and \(\beta\) (slope from Mar to Apr) for nine
randomly selected sites from lilac-honeysuckle dataset (Rosemartin et al. 2015) as measured by 
GHCND stations nearby. Temperatures below 0 are set to 0.}
\label{fig:temp_plot-1}
\end{figure*}

\vfill

\begin{figure*}
\centering
\pandocbounded{\includegraphics[keepaspectratio]{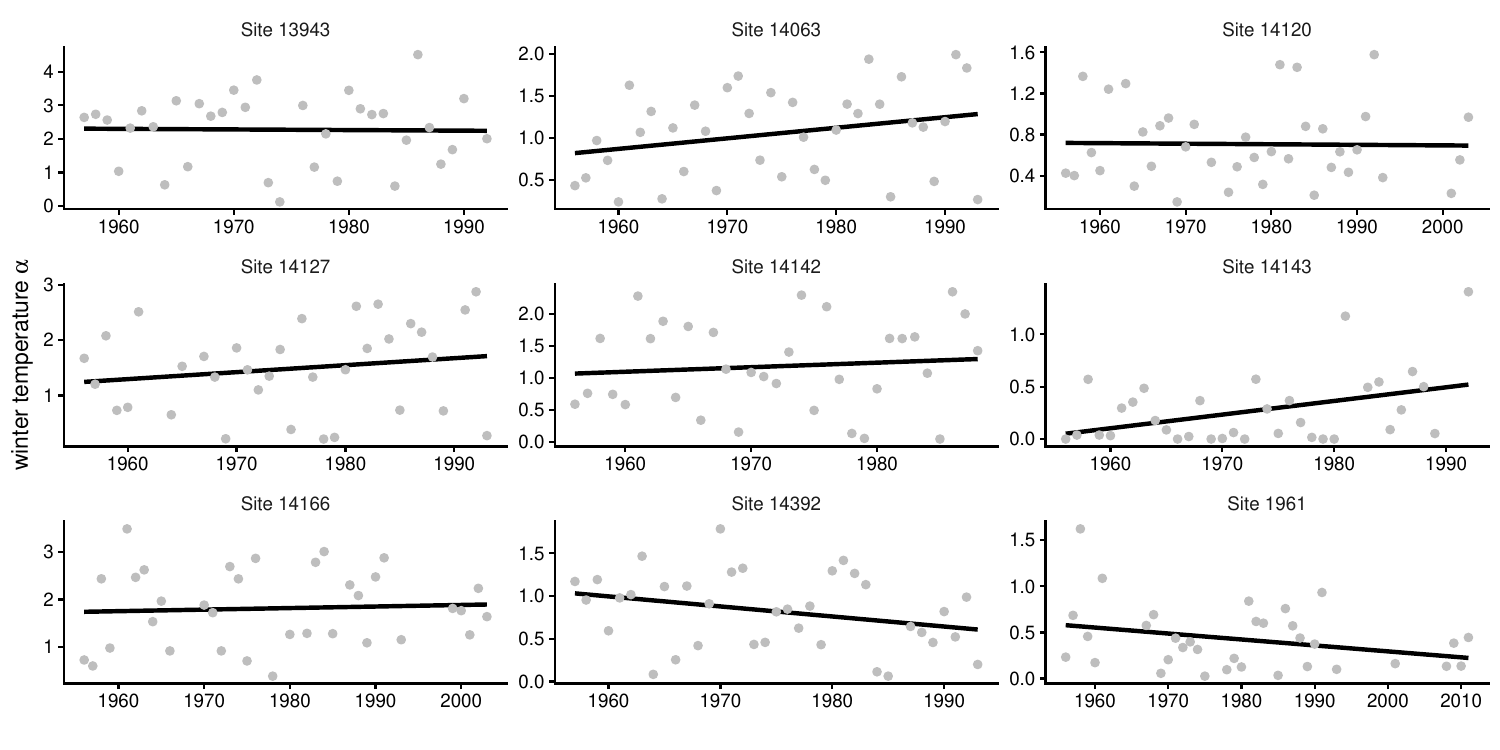}}
\caption{Changes in \(\alpha\) for the nine sites from lilac-honeysuckle data (Rosemartin et al. 2015) with 
longest running records as as measured by GHCND stations nearby. Trend line calculated using linear regression.}
\label{fig:temp_alpha}
\end{figure*}

\begin{figure*}
\centering
\pandocbounded{\includegraphics[keepaspectratio]{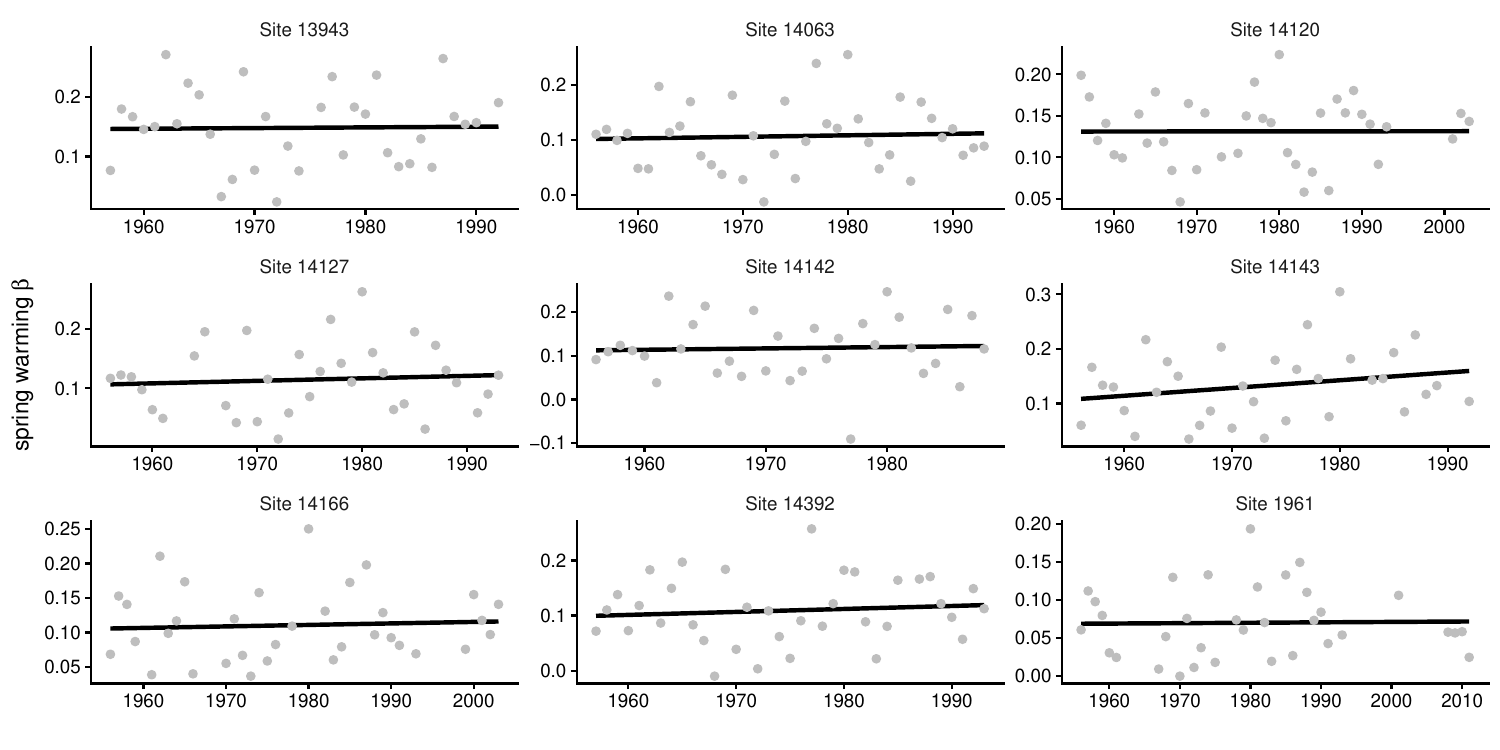}}
\caption{Changes in \(\beta\) for the nine sites from lilac-honeysuckle data (Rosemartin et al. 2015) with 
longest running records as as measured by GHCND stations nearby.  Trend line calculated using linear regression.}
\label{fig:temp_beta}
\end{figure*}

\vfill

\begin{figure*}
\centering
\pandocbounded{\includegraphics[keepaspectratio]{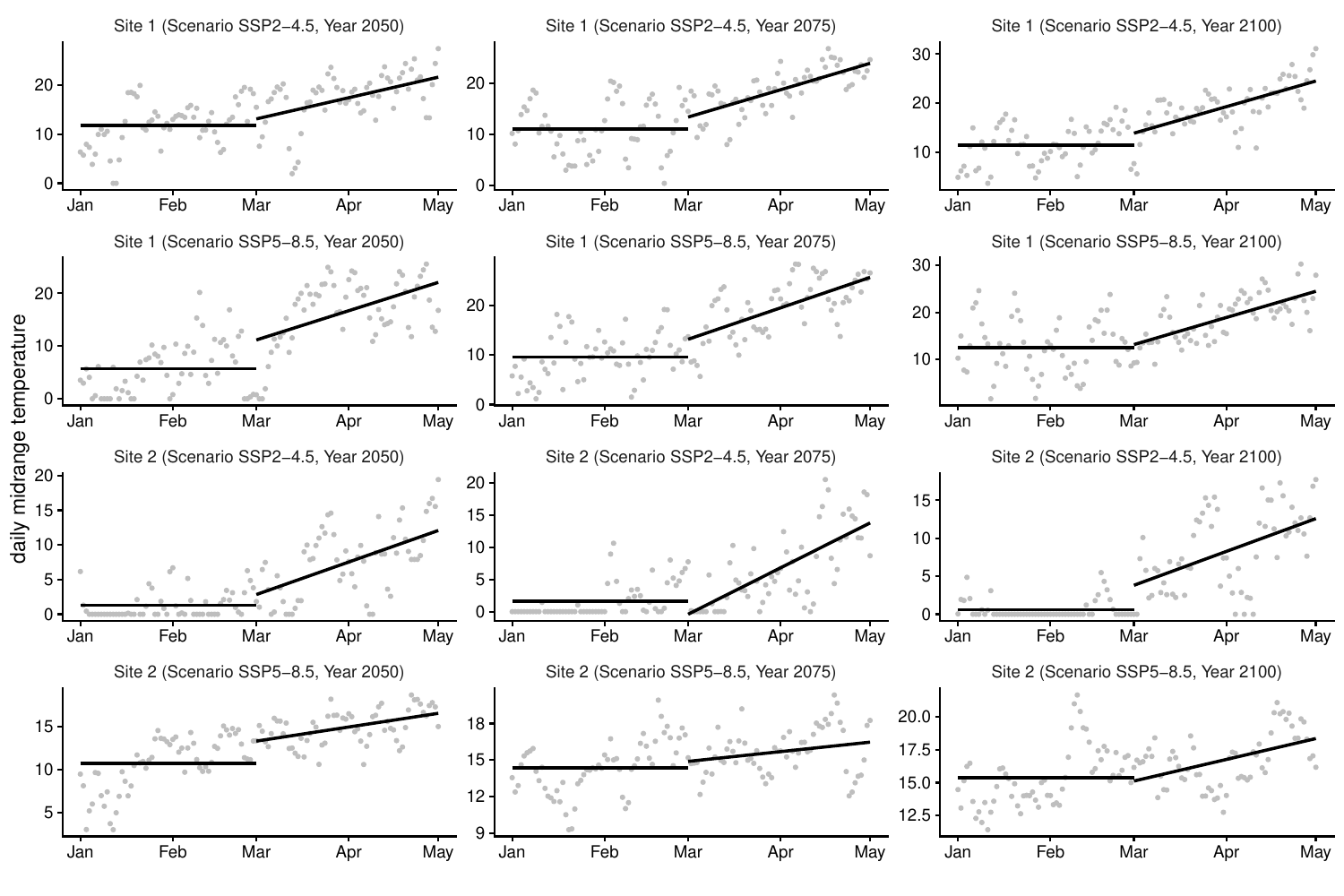}}
\caption{Springtime daily midrange temperatures (points) with \(\alpha\)
(mean from Jan to Feb) and \(\beta\) (slope from Mar to Apr) for two randomly selected sites under two scenarios for the years 2050, 2075, and 20100 from the NEX-GDDP-CMIP6 dataset (Thrasher et al. 2022). Temperatures below 0 are set to 0.}
\label{fig:temp_plot-2}
\end{figure*}

\vfill

\begin{figure*}
\includegraphics[width=1\linewidth]{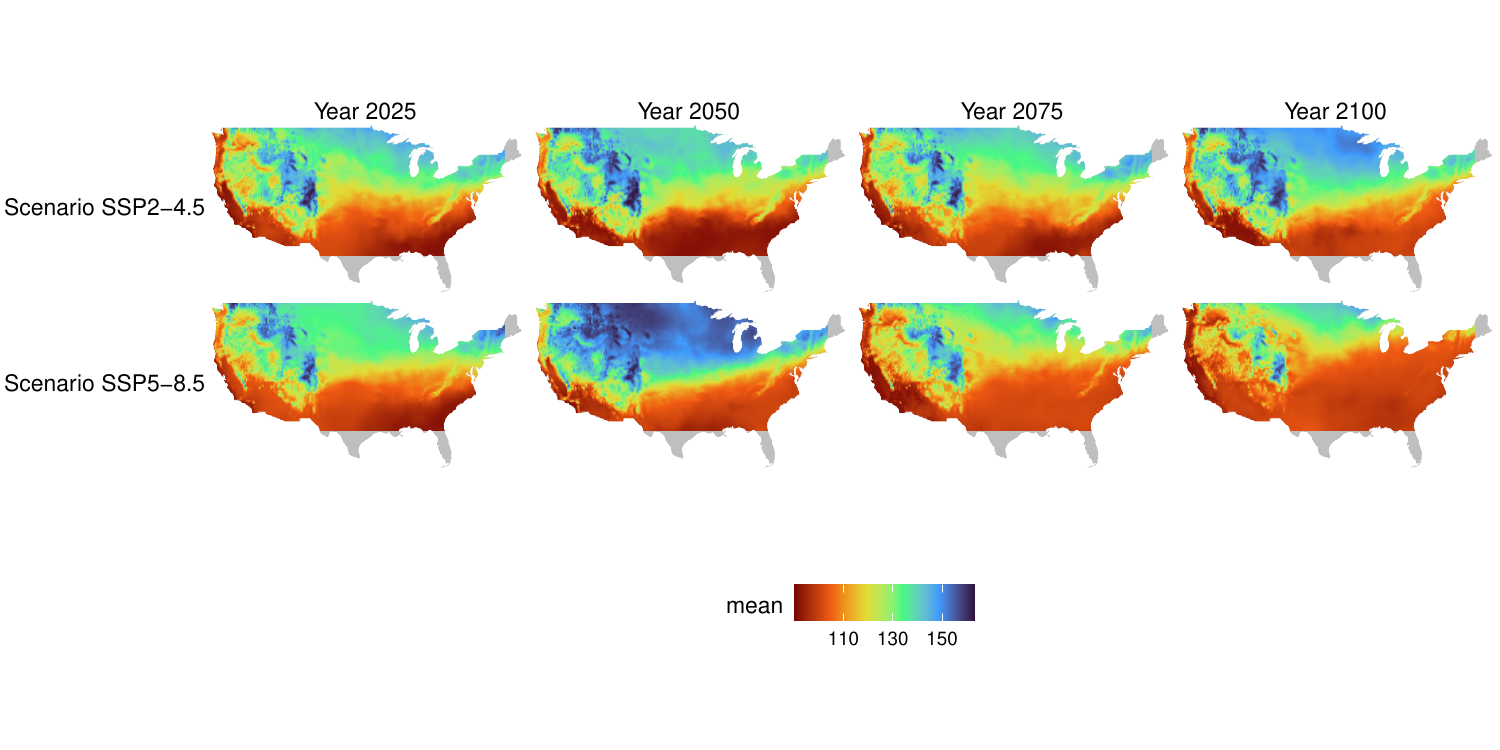} \caption{Predicted mean of lilac bloom dates in years 2025, 2050, 2075, and 2100 (columns) across the continental United States under temperatures projected under two scenarios (rows) from NEX-GDDP-CMIP6 \citep{thrasher2022nasa} in which additional warming is moderate (top) or high (bottom). Lighter areas indicate earlier blooms due to higher temperatures. Under SSP2-4.5, the mean bloom date is largely constant between 2050 and 2100, while under SSP5-8.5, the mean bloom date advances moderately.}
\label{fig:forecast_mean}
\end{figure*}

\begin{figure*}
\includegraphics[width=1\linewidth]{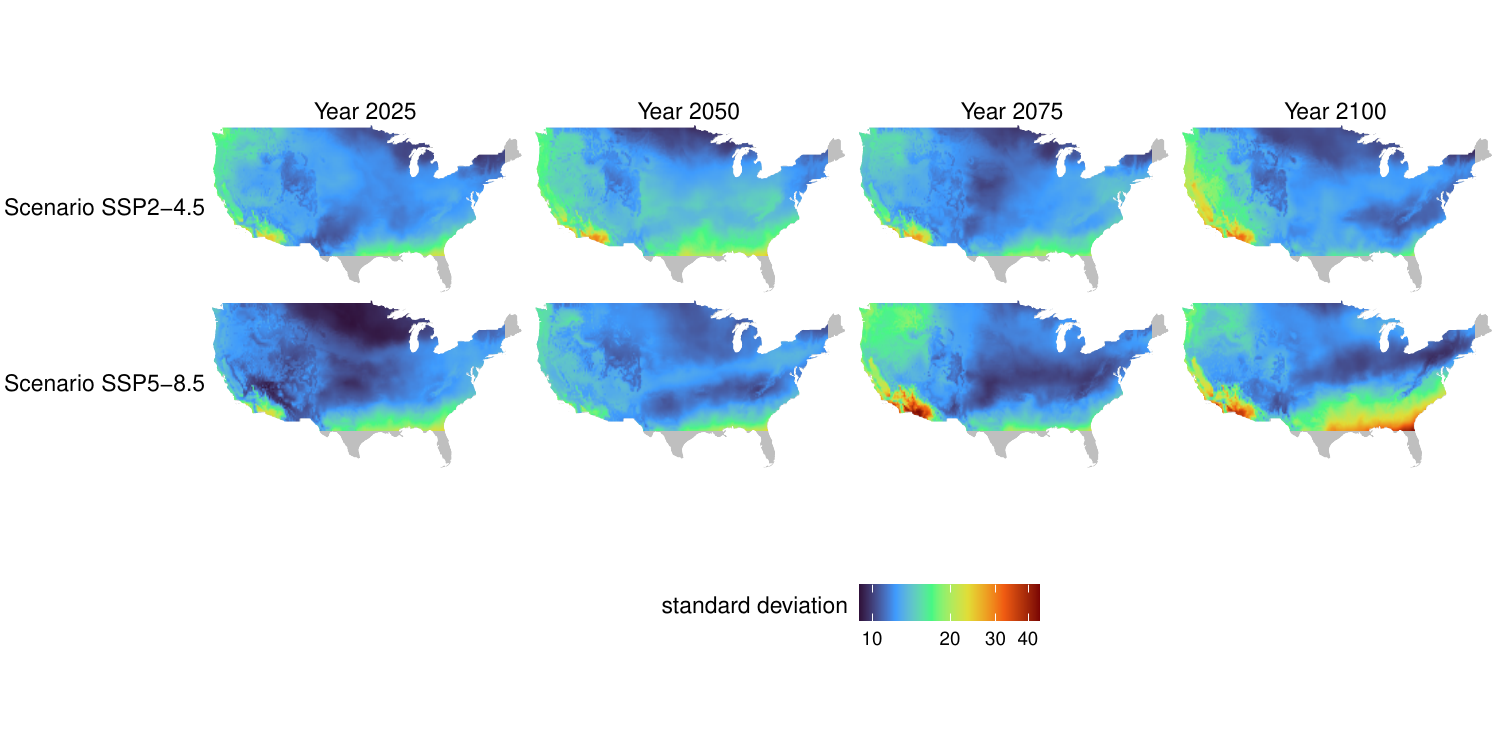} \caption{Predicted standard deviation of lilac bloom dates in years 2025, 2050, 2075, and 2100 (columns) across the continental United States under temperatures projected under two scenarios (rows) from NEX-GDDP-CMIP6 \citep{thrasher2022nasa} in which additional warming is moderate (top) or high (bottom). Light areas indicate more variation (less synchrony) due to higher temperatures. Under SSP2-4.5, the standard deviation is largely constant between 2050 and 2100, while under SSP5-8.5, the standard deviation increases dramatically, particularly in the Southeast and Southwest. These dramatic increase is due to a regime change, which we refer to as a `scattered spring.'}
\label{fig:forecast_var}
\end{figure*}

\newpage

\subsection{SI.4 Simulations}\label{s.4-simulations}

We run two simulations to verify the interpretation of our findings. The
first verifies the asymptotic approximation of the thermal-sum model derived in Section A.1. 
The second verifies the interpretation of the data in Section A.2.

\subsubsection{Simulation 1}\label{simulation-1}

We verify the accuracy of the two asymptotic normal
approximations for the hitting time \[
\nu(\tau)=\min\Big\{n\ge 1:\ \sum_{i=1}^n X_i>\tau\Big\}
\] under the model \[
X_i=\alpha+\beta i+\varepsilon_i,\qquad \varepsilon_i\sim \mathrm{Normal}(0,\sigma^2).
\]

We set \(\sigma=20\) and generate daily temperatures
until the threshold \(\tau\) is reached, recording the corresponding
hitting time \(\nu(\tau)\). We repeat this procedure \(R=10{,}000\) times
for thresholds \(\tau = \{1000,2000\}\), \(\alpha = \{2, 4\}\) and
\(\beta = \{0, 0.1 \}\). When \(\beta = 0\), the simulation matches the
winter regime, and when \(\beta = 0.1\), it matches the spring regime.

We then compare these simulations with the corresponding asymptotic distribution.
When \(\beta=0\), we use the distribution \[
\nu(\tau)\ \dot{\sim}\ \mathrm{Normal}\!\left(\frac{\tau}{\alpha},\ \frac{\sigma^2 \tau}{\alpha^3}\right).
\] When \(\beta>0\), we use the distribution \[
\nu(\tau)\ \dot{\sim}\ \mathrm{Normal}\!\left(\sqrt{\frac{2\tau}{\beta}}-\frac{\alpha}{\beta}+\frac12,\ 
\frac{\sigma^2}{\beta^{3/2}\sqrt{2\tau}}\right).
\]

The simulated distribution of \(\nu(\tau)\) is represented in Figure \ref{fig:simulations}
below by the histograms, where we have standardized by subtracting the mean and dividing by the standard deviation of the 
corresponding asymptotic distribution. Overlaid is the standard normal density. Across both regimes, the standardized 
histograms align closely with the \(\mathrm{Normal}(0,1)\) curve, and the agreement improves as \(\tau\)
increases, providing a visual confirmation of the derived asymptotic
distributions.

\begin{figure*}
\centering
\pandocbounded{\includegraphics[keepaspectratio]{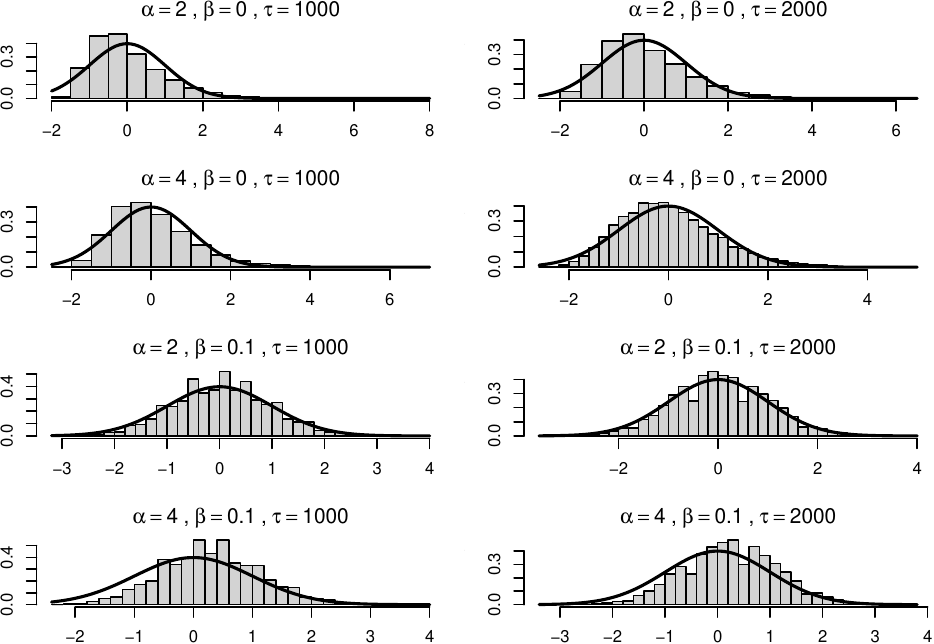}}
\caption{Simulations comparing the distribution of bloom dates (histogram) with an asymptotic approximation (solid line) under two regimes: winter (top two rows) and 
spring (bottom two rows).}
\label{fig:simulations}
\end{figure*}

\subsubsection{Simulation 2}\label{simulation-2}

To verify our interpretation of the data, we simulate daily temperatures \[
X_i=\mu_i+\varepsilon_i,\qquad \varepsilon_i\sim \mathrm{Normal}(0,\sigma^2),
\] with \(\sigma=20\), where the deterministic component follows \[
\mu_i=
\begin{cases}
\alpha, & i=1,\dots,90 \quad \text{(January--February)},\\[2pt]
\alpha+\beta(i-90), & i=91,\dots,180 \quad \text{(March--April)}.
\end{cases}
\] Bloom occurs when cumulative temperature first exceeds a threshold
\(\tau\), i.e. \[
\nu(\tau)=\min\Big\{n\ge 1:\ \sum_{i=1}^n X_i>\tau\Big\}.
\] We simulated \(R=10{,}000\) independent realizations of \(\nu(\tau)\)
for each combination of \(\alpha\in\{4,8,10\}\),
\(\beta\in\{0.2,0.4,0.8\}\), and \(\tau\in\{1000,2000\}\).
We report the mean (left) and standard deviation (right) of \(\nu(\tau)\)
in Tables \ref{tab:sim_mean_1000}-\ref{tab:sim_sd_2000},
arranged with rows indexing \(\alpha\) and columns indexing \(\beta\).

The results reproduce the trends observed in the lilac
data. Mean bloom dates decrease as either
\(\alpha\) increases or \(\beta\) increases
(Tables \ref{tab:sim_mean_1000} and
\ref{tab:sim_mean_2000}). Within a fixed \(\alpha\) row, increasing
\(\beta\) reduces the standard deviation of bloom dates
(Tables \ref{tab:sim_sd_1000} and
\ref{tab:sim_sd_2000}), consistent with the prediction that
increasing average temperatures diminish the influence of noise on the
threshold-crossing time. Holding \(\beta\) fixed, the standard deviation is often
larger at higher \(\alpha\) (particularly for \(\beta\in\{0.4,0.8\}\)),
consistent with the prediction that earlier crossings
occur in a flatter portion of the seasonal cycle where microclimate
variability has greater influence.

\begin{table*}[t]
\centering
\caption{Simulated mean and standard deviation of the bloom date
$\nu(t)$ across $R=10{,}000$ replicates with thresholds $\tau\in\{1000,2000\}$. Rows denote the winter
temperature ($\alpha$), and columns denote spring warming ($\beta$).}
\label{tab:sim_piecewise_all}

\setlength{\tabcolsep}{6pt}
\renewcommand{\arraystretch}{1.15}

\begin{minipage}{0.49\linewidth}
\centering
\subcaption{mean, $\tau=1000$}
\label{tab:sim_mean_1000}
\begin{tabular}{lrrr}
\toprule
$\alpha$ $\backslash$ $\beta$ & 0.2 & 0.4 & 0.8 \\
\midrule
4  & 151.66 & 136.83 & 124.86 \\
8  & 114.87 & 111.27 & 106.85 \\
10 &  98.45 &  97.20 &  95.72 \\
\bottomrule
\end{tabular}

\vspace{1.0em}

\subcaption{mean, $\tau=2000$}
\label{tab:sim_mean_2000}
\begin{tabular}{lrrr}
\toprule
$\alpha$ $\backslash$ $\beta$ & 0.2 & 0.4 & 0.8 \\
\midrule
4  & 199.54 & 171.15 & 149.16 \\
8  & 169.81 & 152.43 & 137.37 \\
10 & 156.22 & 143.40 & 131.34 \\
\bottomrule
\end{tabular}

\end{minipage}
\hfill
\begin{minipage}{0.49\linewidth}
\centering
\subcaption{sd, $\tau=1000$}
\label{tab:sim_sd_1000}
\begin{tabular}{lrrr}
\toprule
$\alpha$ $\backslash$ $\beta$ & 0.2 & 0.4 & 0.8 \\
\midrule
4  & 15.48 & 10.42 &  7.21 \\
8  & 16.53 & 13.27 & 10.50 \\
10 & 15.94 & 14.45 & 12.71 \\
\bottomrule
\end{tabular}
\vspace{1.0em}

\subcaption{sd, $\tau=2000$}
\label{tab:sim_sd_2000}
\begin{tabular}{lrrr}
\toprule
$\alpha$ $\backslash$ $\beta$ & 0.2 & 0.4 & 0.8 \\
\midrule
4  & 10.96 & 7.22 & 4.84 \\
8  & 10.93 & 7.50 & 5.11 \\
10 & 10.69 & 7.66 & 5.36 \\
\bottomrule
\end{tabular}
\end{minipage}
\end{table*}

\end{document}